\documentclass[11pt]{article}
\usepackage{a4,epsfig}
\def\tQ2{\hat{Q}^2}
\def\tq2{\hat{Q}^2}

\renewcommand{\abstract}{}

\begin{document}
{

\begin{flushright}
IFT 2001/13 \\
\end{flushright}

\bigskip
\begin{center}

\section*{Reach of future colliders in probing the structure of the photon}

{\bf M. Krawczyk}\\
Institute of Theoretical Physics, University of Warsaw\\
Warsaw, Poland\\
{\tt krawczyk@fuw.edu.pl}\\
{\bf S. S\"{o}ldner-Rembold}\\
CERN, Geneva\\
Switzerland\\
{\tt Stefan.Soldner-Rembold@cern.ch}\\
{\bf M. Wing}\\
McGill University, Montreal\\
Canada\\
{\tt wing@mail.desy.de}

\end{center}

\begin{center}
\subsubsection*{Abstract}
\end{center}

\abstract{A comparison of the potentials of ep and e$^+$e$^-$ machines to probe the 
structure of the photon is performed. In particular, the kinematic reach of a proposed 
future ep facility, THERA, is compared with those of current colliders, LEP and HERA, 
and with the proposed linear collider, TESLA. THERA like HERA will use a proton beam 
of 920~GeV but with an increased electron beam energy of 250~GeV allowing higher 
scales, $\tq2$, and lower values of parton momentum fraction in the photon, $x_\gamma$, 
to be probed.}

\bigskip

\section{Introduction}

The photon - the gauge boson of QED - has, in high energy processes, a
``hadronic structure''. In deep inelastic scattering, e$\gamma \rightarrow$ e hadrons, 
the corresponding structure function can be introduced, $F_2^{\gamma}(x,\tQ2)$, 
where $\tQ2$ is the scale at which the quasi-real photon is probed. At low Bjorken $x$ this 
structure function 
is expected to behave like $F_2^{\rm p}$, i.e. it increases towards lower $x$ at sufficiently
large $\tq2$, where $\tq2$ is the scale at which the quasi-real photon is probed. 
Unique expectations for the photon are the logarithmic rise of the structure function, 
$F_2^{\gamma}$, with the scale $\tq2$ and a large quark density at large 
$x~(\sim x_{\gamma})$. Observations of these phenomena are basic tests of QCD.

In deep-inelastic e$\gamma$ scattering at e$^+$e$^-$ colliders the scale $\tq2$
is given by the virtuality $Q^2$ of the probing virtual photon whereas in the 
photoproduction regime ($Q^2<1$~GeV$^2$) in ep collisions
and collisions of quasi-real photons, $\gamma \gamma$ scattering, in e$^+$e$^-$ 
colliders it is usually given by the transverse momentum $p_{\rm T}$ of jets 
or final state particles. 

The ep collider, THERA, would offer the opportunity to study the partonic structure of the 
photon extending the kinematic range in $x_\gamma$ and $\tq2$ 
over existing colliders (HERA and LEP) by approximately one order of magnitude.
At lowest order, $x_\gamma$ is equal to unity for ``direct'' processes 
(Fig.~\ref{php_feynman}a), whereas ``resolved'' processes (Fig.~\ref{php_feynman}a), where 
the photon interacts via its partonic content, are characterised by a smaller 
value of $x_\gamma$. In addition, the photoproduction of particles (hadrons or prompt photons)
or jets at high $p_{\rm T}$ provides complementary information to that from 
deep-inelastic e$\gamma$ on the partonic, in particular gluonic 
(Fig.~\ref{php_feynman}b), content of the quasi-real photon. The photoproduction of 
dijets, heavy quarks and prompt photons have been studied~\cite{jets-heavy-jkw,azmk-prompt}, 
with the emphasis on the potential of THERA to yield information on the structure of the 
real photon. The possibility of studying the structure of the virtual photon at THERA has 
also been considered~\cite{ujd-long}. 

\begin{figure}[htp]
\unitlength=1mm                      
\begin{picture}(0,0)(100,100)
\put(142,40){\bf \Large{(a)}}
\put(208,40){\bf \Large{(b)}}
\end{picture}
\begin{center}
~\epsfig{file=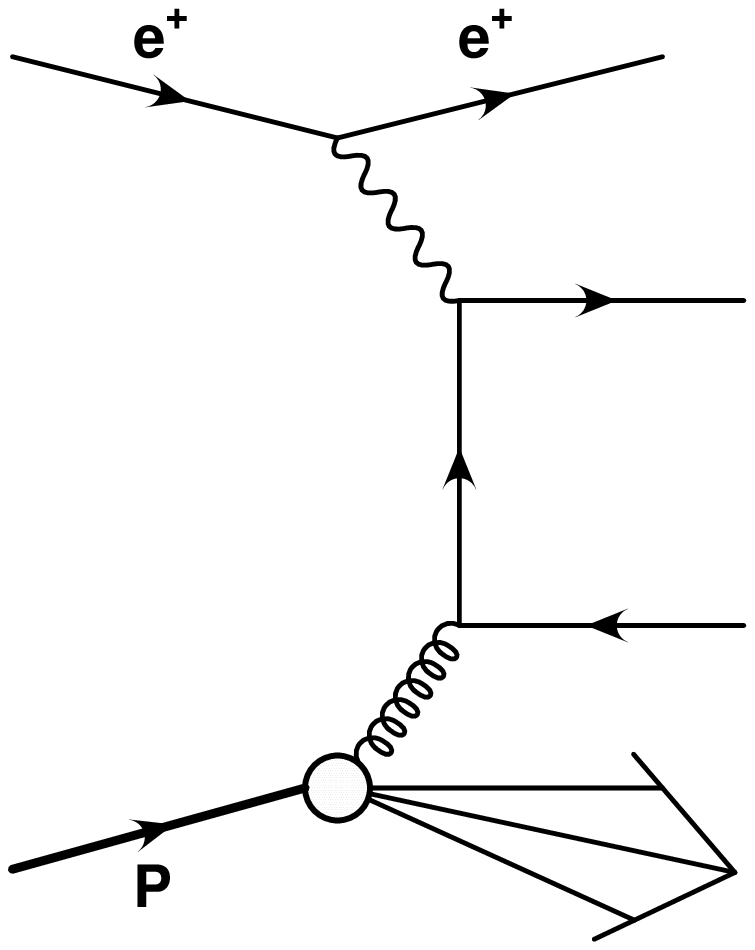,height=5.5cm}
\hspace{2cm}~\epsfig{file=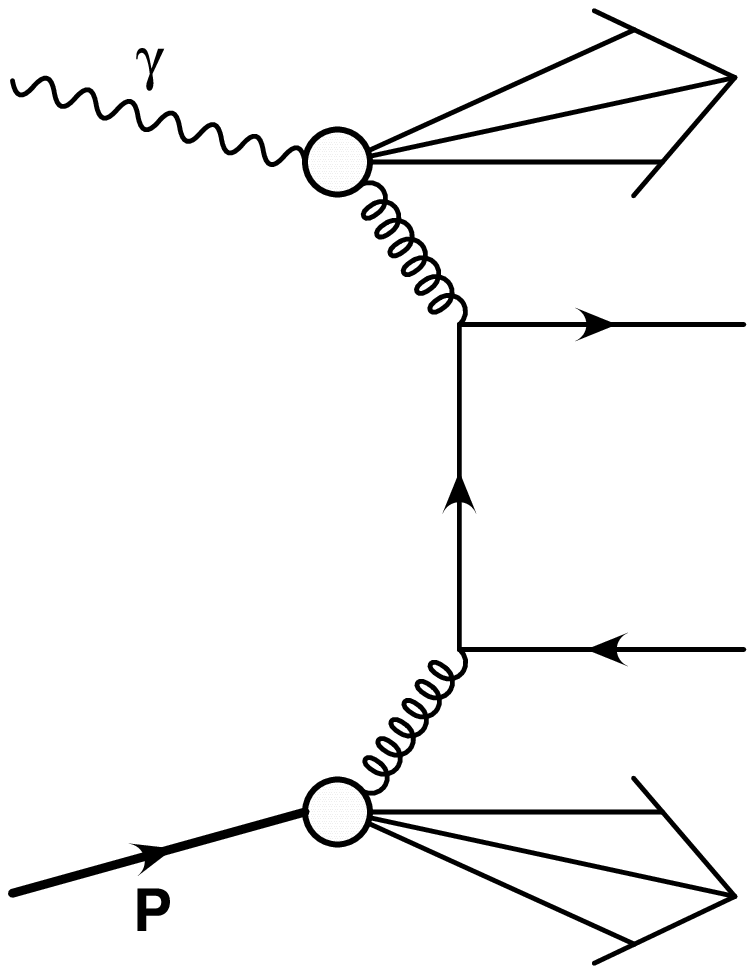,height=5.5cm}
\end{center}
\caption{Examples of lowest order (a) direct photon and (b) resolved photon processes in ep 
collisions.}
\label{php_feynman}
\end{figure}

Good knowledge of the hadronic interactions of a fundamental particle - the photon - 
is essential for the future high energy physics programme. The present situation is 
not satisfactory as data for some processes, such as that for the photoproduction of 
dijets at HERA are not in agreement with existing next-to-leading (NLO) QCD 
calculations~\cite{zeus_highet,general}. The agreement for processes involving resolved 
virtual photons is even more problematic. A proper description of the hadronic interaction 
of photons is also needed to calculate the Standard Model background in searches for the 
Higgs particle and other new phenomena at future colliders.

Initially a comparison is made of the kinematic regions accessible at various 
colliders where the structure of the photon can be tested: LEP, HERA and 
the future linear collider (TESLA) in the basic e$^+$e$^-$ mode. 
The aim of this section is to show how THERA can enrich the potential of TESLA 
in its standard e$^+$e$^-$ mode and so detailed comparisons with photon colliders 
($\gamma \gamma$ or e$\gamma$ modes) 
are not considered. It is assumed that the energy of the electron beam is 250 GeV 
and that of the proton beam is 920 GeV.

\section{The THERA kinematic region in comparison with other colliders}

In considering the benefits of THERA for studying the structure of the photon, all current 
machines and those of the future are discussed for comparison. The kinematic reach 
of THERA ($\sqrt{s}\approx 1$ TeV) is compared with those of LEP 
($\sqrt{s} \approx 200$ GeV), HERA ($\sqrt{s}\approx 300$ GeV) and a future 
linear  e$^+$e$^-$ collider, TESLA, ($\sqrt{s}\approx 500$ GeV), where $\sqrt{s}$ 
denotes the centre-of-mass energy of the colliding primary beams. The nominal ep 
and e$^+$e$^-$ options are considered for THERA and TESLA, respectively; the corresponding 
$\gamma$p and $\gamma$e or $\gamma \gamma$ options cover a similar kinematic region with 
an increased cross section but lower luminosity. Of interest is to consider the 
minimum $x_\gamma$, the range of $\tq2$ and the polar angle (rapidity) of the jets in 
resolved photon events. For comparison the following quantities are introduced, relevant 
for deep-inelastic e$\gamma$ scattering and resolved photon processes in e$^+$e$^-$ and 
ep colliders:

\begin{equation}
Q^2_{\rm max,min}|_{\rm e^+e^-} = \frac{W^2_{\rm max,min} x_\gamma}{1-x_\gamma}, \ \ \ \ \ \ 
\tq2|_{\rm ep} \equiv p_{\rm T}^2 = \left( y_{\rm max} E_{\rm e} e^{\eta _{\rm max}} x_\gamma \right)^2
\end{equation}

\begin{equation}
x|_{\rm e^+e^-}=\frac{Q^2}{Q^2+W^2},\ \ \ \
x_\gamma^{\rm min}|_{\rm e^+e^-} = \frac{p_{\rm T} e^{\pm \eta_{\rm CM}}}
                                  {2 E_{\rm e} - p_{\rm T} e^{\mp \eta_{\rm CM}}}, \ \ \ \ \ \ 
x_\gamma^{\rm min}|_{\rm ep} = \frac{E_{\rm p} p_{\rm T} e^{-\eta_{\rm LAB}}}
                              {2 E_{\rm e} E_{\rm p} - E_{\rm e} p_{\rm T} e^{\eta_{\rm LAB}}},
\end{equation}
where $W$ denotes the invariant mass of the hadronic final state.
In Fig.~\ref{kinematics_php}a, the minimum photon momentum fraction, 
$x_\gamma^{\rm min}$, is shown for a given transverse momentum, $p_{\rm T}, (=E_{\rm T}$ for 
massless particles) of 10 GeV, as a function of the rapidity, $\eta$, of the jets in the 
laboratory frame for e$^+$e$^-$ colliders (equivalent to the centre-of mass frame) 
and ep colliders. It can be seen that for a given rapidity, an order of 
magnitude smaller value of $x_\gamma^{\rm min}$ at THERA can be probed compared 
with HERA due to the increased electron beam energy, $E_{\rm e}$. The minimum $x_\gamma^{\rm min}$ 
at TESLA would also extend the minimum possible at LEP and HERA. However, smaller 
values of $x_\gamma^{\rm min}$ can be reached at THERA than at TESLA in the very 
forward rapidity direction ($\eta^{\rm ep}_{\rm LAB}>2$) reaching a minimum for this 
transverse momentum at $\eta^{\rm ep}_{\rm LAB}\sim4.6$. This demonstrates that good 
forward detectors are needed for THERA with the ability to accurately 
reconstruct jets up to the rapidities discussed here. Only the e$^+$e$^-$ TESLA collider 
is considered. The kinematic reach of the e$^+$e$^-$ TESLA collider depends very much on the 
minimum required for the energy $E_{\rm tag}$ and angle $\theta_{\rm tag}$ of the scattered 
electron. A much larger kinematic range could be covered with an e$\gamma$ collider based on 
TESLA~\cite{vogt}

\begin{figure}[htp]
\unitlength=1mm
\begin{picture}(0,0)(100,100)
\put(142,87){\bf \Large{(a)}}
\put(220,87){\bf \Large{(b)}}
\end{picture}
\begin{center}
~\epsfig{file=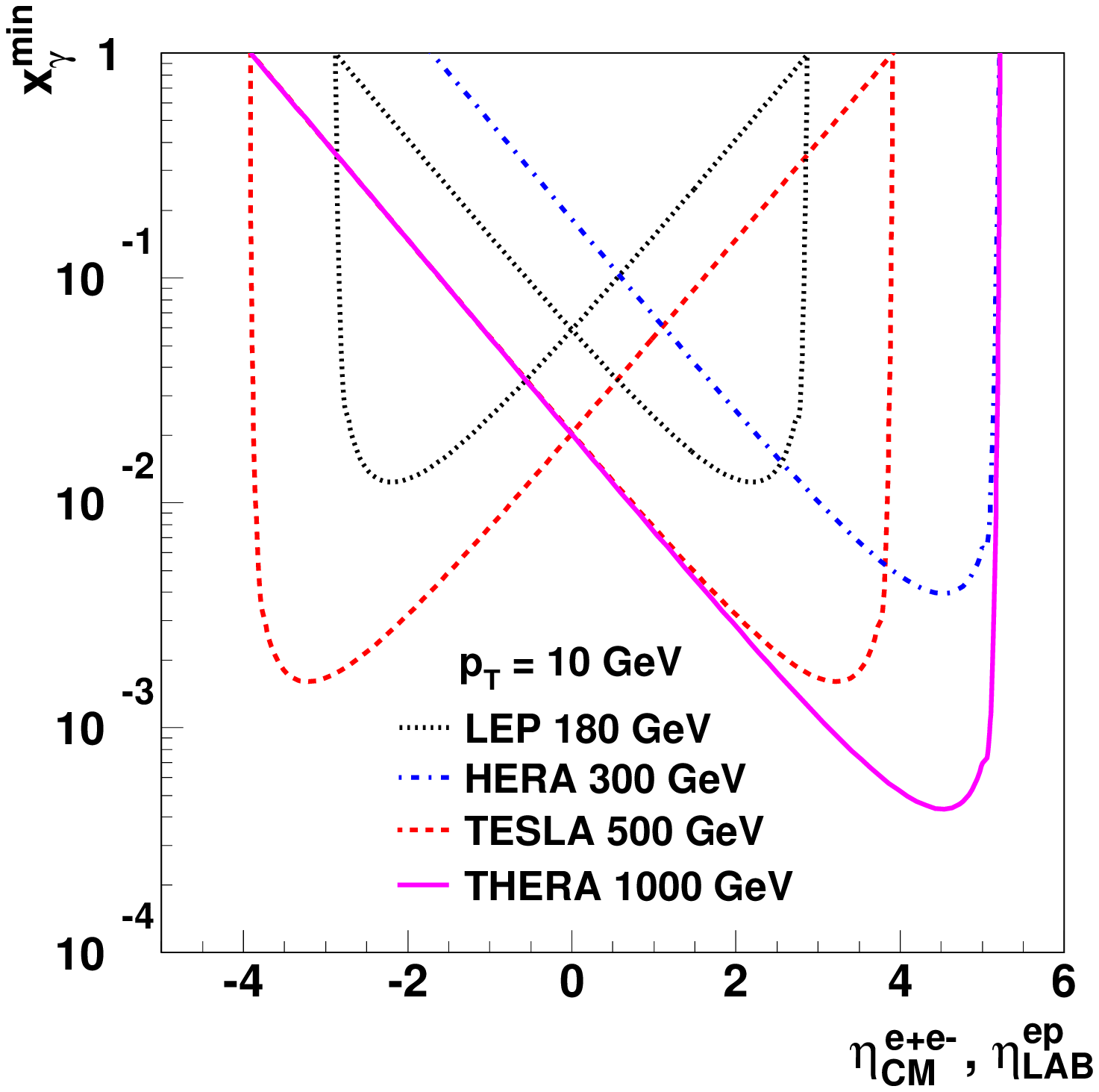,height=7.7cm}
~\epsfig{file=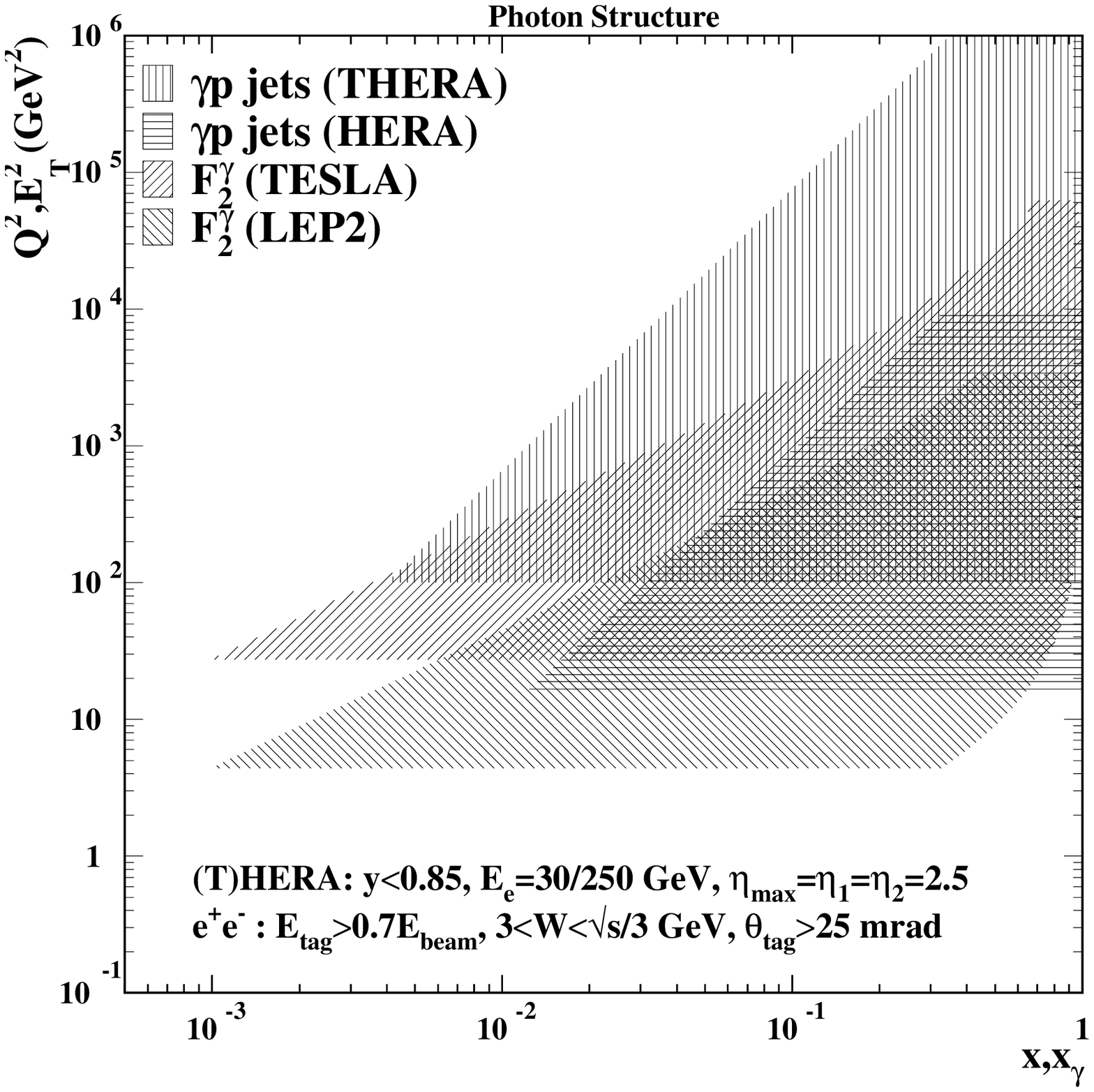,height=7.7cm}
\end{center}
\vspace{-0.5cm}
\caption{(a) The minimum photon momentum fraction, $x_\gamma^{\rm min}$, versus the 
rapidity of the jets in the centre-of-mass frame for e$^+$e$^-$ machines and laboratory frame for 
ep machines. (b) Range in $\tq2$ ($Q^2$ for $F_2^\gamma$ or $p_{\rm T}^2$ for $\gamma$p jets) 
versus $x_\gamma$ with realistic 
detector scenarios shown. The region for THERA is compared with that of 
TESLA, HERA and LEP.}
\label{kinematics_php}
\end{figure}

Considering some restrictions in the detector layout, the values of $\tq2$ obtainable 
are shown versus $x_\gamma$ in Fig.~\ref{kinematics_php}b. Detectable scenarios for 
LEP and HERA are described in Fig.~\ref{kinematics_php}b and the same 
for TESLA and THERA are also imposed, although it is hoped that the future experiments   
would have improved detectors in the very forward and backward regions. Here it can 
be seen that although the e$^+$e$^-$ machines will yield the lowest values of 
$x_\gamma$, it is also apparent that the ep machines can probe a smaller value 
of $x_\gamma$ for a given $\tq2$. In particular, THERA will provide valuable 
additional information on the structure of the photon in the region, $x_\gamma>0.01$, 
particularly at high-$p_{\rm T}$, complementing TESLA and the current experiments.

\section{Summary}

Photoproduction at THERA can further current knowledge of the structure of the photon, 
extending the current colliders, HERA and LEP and complementing the future linear e$^+$e$^-$
collider program. The kinematic range can be extended, for quasi-real photons,  in $x_\gamma$ 
and the hard scale, $\tq2 \sim p_{\rm T}^2$. Also the structure of the virtual photon for 
larger square of its mass, $Q^2$, can be probed. Inclusive dijets, heavy quarks and prompt 
photons have been 
studied as tools to probe the structure of the quasi-real and virtual photon. The 
building of THERA will, therefore, enrich the field on the structure of a fundamental 
gauge boson - the photon.

\section*{Acknowledgements}

We would like to thank the organisers for an interesting and productive series of 
workshops. One of us (MK) have been partly supported by the Polish State Committee
for Scientific Research (grant 2P03B05119, 2000-2001). MK is also grateful to the 
European Commission 50th Framework Contract HPRN-CT-2000-00149 and DESY for financial 
support.

}
\end{document}